\documentclass[sigconf]{acmart}

\renewcommand\footnotetextcopyrightpermission[1]{} % removes footnote with conference information in first column
\usepackage{caption,subcaption}

\AtBeginDocument{%
  \providecommand\BibTeX{{%
    \normalfont B\kern-0.5em{\scshape i\kern-0.25em b}\kern-0.8em\TeX}}}

\acmConference[arXiv '20]{arXiv '20: IBM Research, India}
\acmBooktitle{IBM Research, India}
\begin{document}

\title{Data Augmentation for Personal Knowledge Base Population}

\author{Lingraj S Vannur}
\affiliation{%
  \institution{Nitte Meenakshi Institute of Technology}
  \city{Bengaluru}
  \country{India}
}
\email{lingrajsv1998@gmail.com}
\author{Balaji Ganesan}
\affiliation{%
  \institution{IBM Research India}
  \city{Bengaluru}
  \country{India}
}
\email{bganesa1@in.ibm.com}
\author{Lokesh Nagalapatti}
\affiliation{%
  \institution{IBM Research India}
  \city{Bengaluru}
  \country{India}
}
\email{lokesn21@in.ibm.com}
\author{Hima Patel}
\affiliation{%
  \institution{IBM Research India}
  \city{Bengaluru}
  \country{India}
}
\email{himapatel@in.ibm.com}
\author{MN Thippeswamy}
\affiliation{%
  \institution{Nitte Meenakshi Institute of Technology}
  \city{Bengaluru}
  \country{India}
}
\email{thippeswamy.mn@nmit.ac.in}

\renewcommand{\shortauthors}{Vannur and Ganesan, et al.}

\begin{abstract}
 Cold start knowledge base population (KBP) is the problem of populating a knowledge base from unstructured documents. While artificial neural networks have led to significant improvements in the different tasks that are part of KBP, the overall F1 of the end-to-end system remains quite low. This problem is more acute in personal knowledge bases, which present additional challenges with regard to data protection, fairness and privacy. In this work, we present a system that uses rule based annotators and a graph neural network for missing link prediction, to populate a more complete, fair and diverse knowledge base from the TACRED dataset.
\end{abstract}




%  \includegraphics[width=\textwidth]{teaser_image.png}
%  \caption{Extracting personal data from text}
%  \Description{A sample text with entities and relations extracted.}

\maketitle

\section{Introduction}
The NIST TAC Knowledge Base Population (KBP) challenges introduced the cold start knowledge base population problem. It involves several tasks that go into populating a knowledge base from unstructured documents. Some of the common tasks include Entity Recognition, Entity Classification, Entity Resolution, Relation Extraction, and Slot Filling. As \cite{zhang2017position} showed while introducing the TACRED dataset, the problem remains largely unsolved with the F1 of the overall relation and slot filling system achieving only 26.7\% F1, which is not sufficient for real world applications.

There have been efforts in the recent past to understand the low overall F1. \cite{alt2020tacred} investigated the harder samples in the TACRED Dataset and concluded that some of the labels need to be re-labeled. Our work examines the TACRED dataset from a different perspective. We investigate the diversity of the protected attributes like gender, location and religion in the populated Knowledge Base so that this cold start approach can be used for Personal Knowledge Base Population.

\begin{figure}[htb]
    \includegraphics[width=0.7\columnwidth]{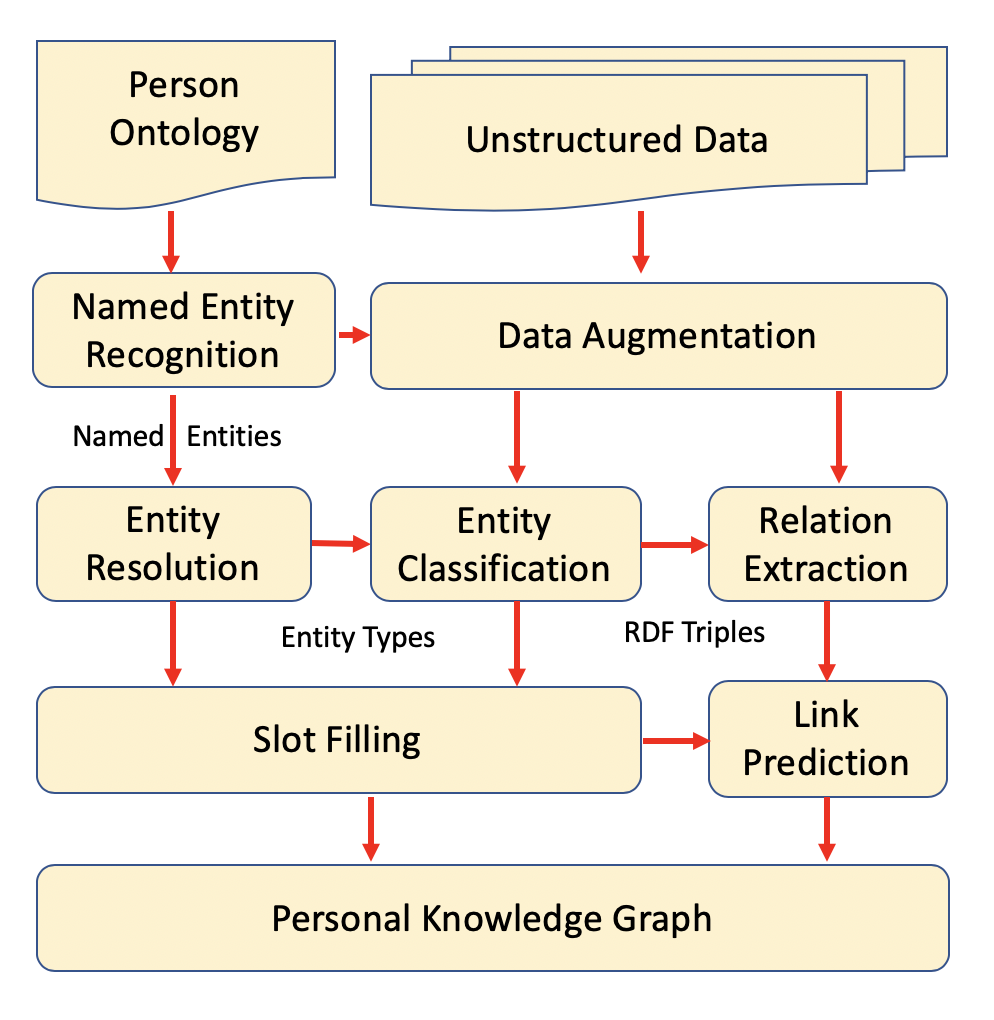}
    \caption{Personal Knowledge Base Population pipeline}
    \label{fig:pipeline}
\end{figure}

\begin{figure*}
    \centering
    \includegraphics[width=\linewidth]{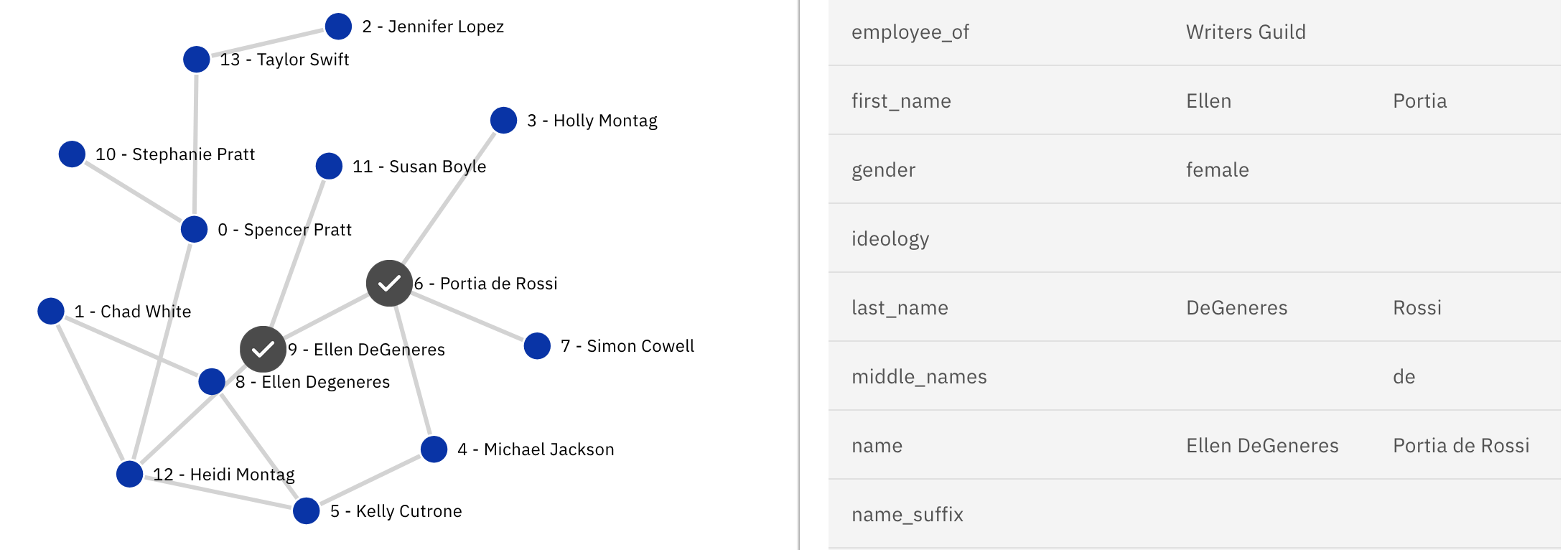}
    \caption{Our goal is to populate a Personal Knowledge Base from TACRED dataset}
    \label{fig:link_prediction}
\end{figure*}

For a number of applications in Data Protection, fraud prevention and business intelligence, there is a need to extract Personal Data Entities, classify them at a fine grained level, and identify relationships between people. Manually created Personal Ontologies and Knowledge Bases are predominantly used for this purpose.

In the above enterprise applications, we typically have Property Graphs where people are nodes, relations between people in the real world are edges, and other kinds of relations become attributes of the nodes. Constructing such a property graph of people, from unstructured documents like emails in an organization, government data of citizens, personal conversations etc is a significantly harder problem than populating general purpose world knowledge bases.

The first challenge in personal KBP is in identifying attributes at a fine grained level. For example, \textit{Brigham Young University} could be classified coarsely as ORGANISATION by a Named Entity Recognizer. In recent years, a number of Neural Fine Grained Entity Classification (NFGEC) models have been proposed, which assign fine grained labels to entities based on context. They could type \textit{Brigham Young University} as \slash{org}\slash{education}. However the focus of such systems has not been on PDEs. They do not treat the problem of identifying PDEs any different from other entities. For personal KBP, we want the entity classification model to assign \textit{educated\_at} as the class for \textit{Brigham Young University}.

Then there are further challenges in relation extraction and missing link prediction. Graph Neural Networks have recently been used for predicting links between people, by representing a node in terms of attributes, edges and the node representations of neighbors and other nodes. GNNs are typically considered black box models and need explanations before they can be used to predict links between people in the real world.

\subsection{Privacy, fairness, and Data Protection}

Considering people knowledge graphs need to have sufficient personal information to be of any practical use, they also have to protect the privacy of the people concerned. So such Knowledge Base population typically happens in-house and remains a largely manual process till now. 

Even within an organization, populating personal knowledge bases can only happen with the informed consent of the employees. Customer data of companies cannot be used for populating such graphs because of privacy reasons and also because Data Protection laws like General Data Protection Regulation (GDPR) forbid such usage of data without the consent of the customers. Although, customers of a company may also benefit from such a personal knowledge graph, to serve Data Subject Access Requests (DSAR) for example, obtaining informed consent from customers will be very hard.

While populating such a knowledge base with personal data, we have to strive to avoid bias in the training data on gender, age, ethnicity, location, religion, sexual orientation among other attributes. While there are methods available to detect bias in models, eliminating bias in the training data can have more desirable outcomes. Business Intelligence and other applications will also benefit from eliminating or reducing bias in the data. Hence in this work, we present ways to augment the training data of neural models used for knowledge base population, as well as for using the populated knowledge base for downstream applications.

\subsection{Data Augmentation}

Data Augmentation is the process of increasing the diversity in the training data without necessarily having to acquire more data. In this work, we show how the popular TACRED Dataset can be made more diverse by annotating more entities in the sentences provided with the dataset. Our motivation, as explained earlier, is to have more features to train a Graph Neural Networks, while simultaneously increase the overall diversity of the populated Personal Knowledge Base. 

We summarize our contributions in this work as follows:

\begin{itemize}
    \item We show the overall increase in diversity and F1 in the Knowledge Base Population task, because of our data augmentation strategies.
    \item Using a post-hoc interpretable model and techniques like LIME and SHAP, we show that the augmented dataset can be used to train Graph Neural Networks, with the confidence that the predictions are not biased against any minority groups.
    \item Finally, we present a representative sampling technique which further reinforces that data augmentation increases the diversity of the populated knowledge base.
\end{itemize}
    
\section{Related Work}

Knowledge Base Population (KBP) has been a fairly well researched problem. The KBP Track at TAC \cite{ji2010overview} that were held between 2010 and 2017 led to significant advances in KBP. Recently \cite{mesquita2019knowledgenet} have introduced a benchmark dataset for the KBP problem. In this work though we focus more on the Personal Data in the Knowledge Base, given our goal to make the KBP process fair.

\cite{balog2019personal} introduced the concept of Personal Knowledge Graphs. A typical pipeline to populate a knowledge graph comprises of Entity Recognition, Entity Classification, Entity Resolution, Slot Filling, Relation Extraction and Link Prediction. \cite{li2014personal} introduced personal knowledge graphs in the context of conversational systems. In this paper, we describe a system to extract personal knowledge graph from the TACRED relation extraction dataset. We treat this as an acceptable proxy for real world enterprise documents like emails, internal wiki pages, organization charts which cannot be used for research because of privacy reasons.

Entity classification is a well known research problem in Natural Language Processing (NLP).~\cite{ling2012fine} proposed the FIGER system for fine grained entity recognition. ~\cite{yogatama2015embedding} showed the relevance of hand-crafted features for entity classification.~ \cite{shimaoka2017neural} further showed that entity classification performance varies significantly based on the input dataset (more than usually expected in other NLP tasks). \cite{dasgupta2018fine} and \cite{abhishek2018collective} introduced model improvements to achieve better results on OntoNotes and Wiki datasets respectively. \cite{ganesan2020neural} proposed a neural architecture to populate a Person Ontology.

Relation Extraction is again a well known research problem. \cite{zhang2017position} who proposed the TACRED dataset also introduced an LSTM sequence model with entity position-aware attention for the task. They later showed in \cite{zhang2018graph} that graph convolution over pruned dependency trees improves relation extraction. More recently \cite{alt2020tacred} revisited the TACRED dataset to understand why state of the art models still struggle with a ceiling on their performance. They report that many of the labels may need to be relabeled. 

\cite{min2020syntactic} showed how pre-trained language models could have better inference performance with syntactic data augmentation.

\cite{bellamy2018ai} introduced a AI Fairness toolkit called AIF360 which provides a number of algorithms for detecting and mitigating bias against protected attributes like gender, age, ethnicity, location, sexual orientation and few others. \cite{garg2019counterfactual} proposed a method using counter factual data to improve fairness. 

%\begin{figure*}
%    \centering
%    \includegraphics[width=\linewidth]{attributes_and_relations.png}
%    \caption{Attributes and Relations}
%    \label{fig:pde_hierarchy}
%\end{figure*}

\section{Data Augmentation}

\begin{figure}
    \centering
    \includegraphics[width=0.8\columnwidth]{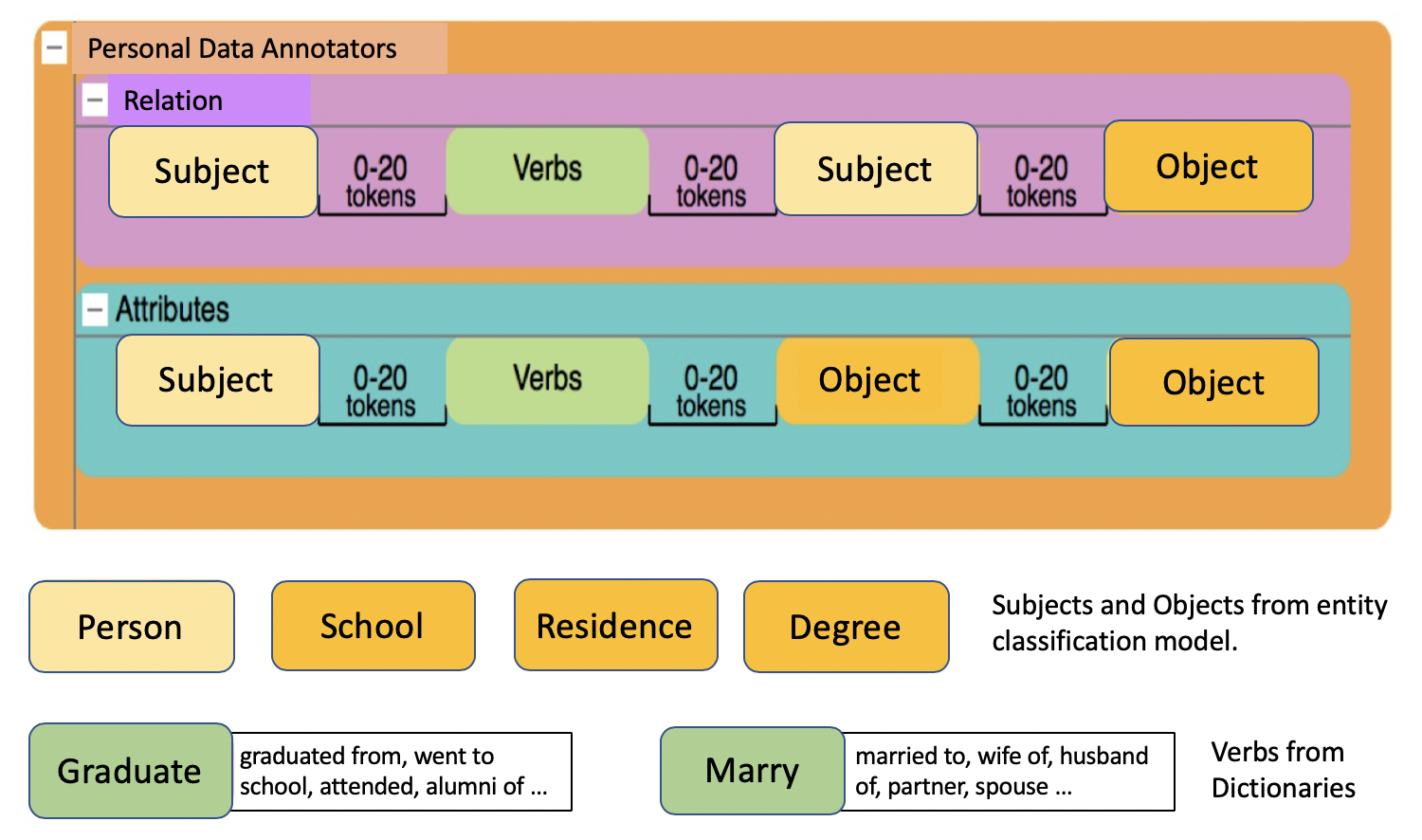}
    \caption{Personal Data Annotators}
    \label{fig:relation_annotators}
\end{figure}

\begin{figure}
    \centering
    \includegraphics[width=\columnwidth]{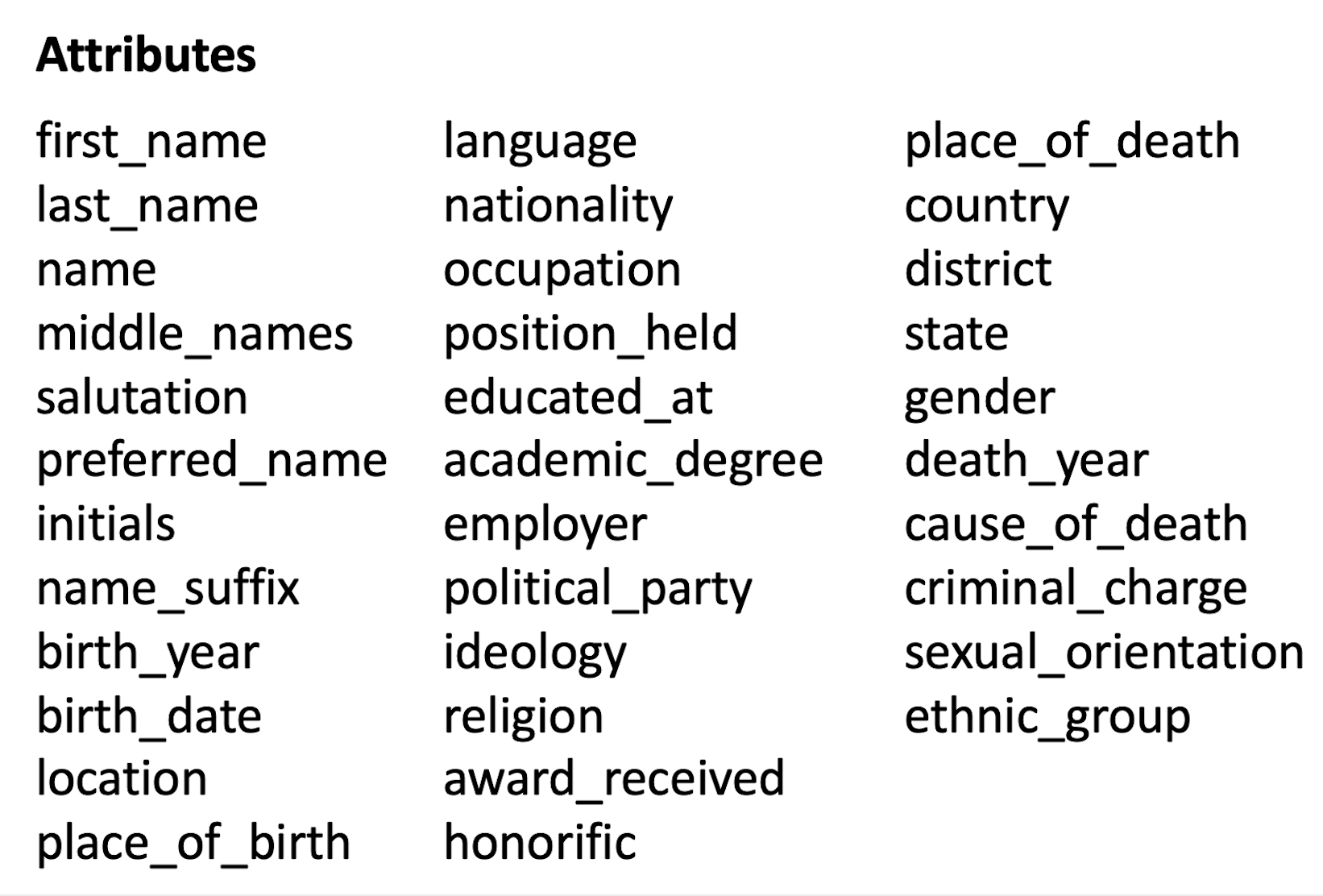}
    \caption{OntoNotes Entity Types}
    \label{fig:ontonotes_entity_types}
\end{figure}

\cite{ling2012fine} proposed the FIGER entity type hierarchy with 112 types.~\cite{gillick2014context} proposed the Google Fine Type (GFT) hierarchy and annotated 12,017 entity mentions with a total of 89 types from their label set. These two hierarchies are general purpose labels covering a wide variety of domains. \cite{dasgupta2018fine} proposed a larger set of Personal Data Entity Types with 134 entity types. We have selected the 34 personal data entity types, as shown in Figure \ref{fig:ontonotes_entity_types} that were found in our input corpus.

For relation extraction labelset, YAGO \cite{yago} contained 17 relations, TACRED \cite{zhang2017position} proposed 41 relations and UDBMS (DBPedia Person) dataset \cite{UDMS_dataset} proposed 9 relations.

\subsection{Personal Data Annotators}

Any system that assigns a label to a span of text can be called an annotator. In our case, these annotators assign an entity type to every entity mention. We have experimented with an enterprise (rule/pattern based) annotation system called SystemT introduced by \cite{chiticariu2010systemt}. SystemT provides about 25 labels, which are predominantly coarse grained labels.

We use these SystemT annotators in 3 ways:
\begin{itemize}
    \item To annotate unstructured data with entities for the entity classification task and augment the output of the classification model.
    \item To provide coarse grained entity types as features to the model described in \cite{dasgupta2018fine}.
    \item To annotate unstructured data with relations for the relation extraction task and augment the output of the relation extraction model.
\end{itemize}

While neural networks have recently improved the performance of entity classification on general entity mentions, pattern matching and dictionary based systems continue to be used for identifying personal data entities in the industry. We believe our proposed approach, consisting of modifications to state-of-the-art neural networks, will work on personal datasets for two reasons.~\cite{yogatama2015embedding} showed that hand-crafted features help, and~\cite{shimaoka2017neural} have shown that performance varies based on training data domain. We have incorporated these observations into our model, by using coarse types from rule-based annotators as side information. 

We used our Personal Data Annotators to create a number of labeling functions like those shown in Figure \ref{fig:relation_annotators} to create a set of relations between the entities. We have created this dataset from the sentences in the TACRED dataset. We ran the Personal Data Annotators on these sentences, providing the bulk of the annotations.

%that are reported in Table~\ref{tab:datasets}. TODO - add a table with additional annotations given by SystemT

We use the method described in~\cite{chiticariu2010systemt} to identify the span of the entity mentions. This method requires creation of dictionaries each named after the entity type, and populated with entity mentions. This approach does not take the context of the entity mentions while assigning labels and hence the data is somewhat noisy. However, labels for name, email address, location, website do not suffer much from the lack of context and hence we go ahead and annotate them.

\section{Experiments}

We have implemented a pipeline for Personal Knowledge Graph population as shown in Figure \ref{fig:pipeline}. This pipeline consists of existing personal data annotators, Stanford Named Entity Recognizer which provide rule based entity and relation extraction. We have then improved two state of the art models for entity classification and relation extraction as described in the previous sections. Finally we use a graph neural network for Link Prediction to infer more relationships between people mentioned in the corpus. The use of a GNN for Link Prediction is to leverage the attributes of the nodes, along with neighboring nodes and edges. While we are doing entity resolution in this task, we use a system similar to the SystemER introduced by \cite{qian2019systemer}.

The input to our pipeline are sentences. The outputs are person entities, their personal data as attributes and semantically rich relations between person entities. These can be used to populate a graph database like the one provided by networkx \cite{hagberg2008exploring}.
\subsection{Entity Classification}

We use the neural network model from~\cite{dasgupta2018fine}, which consists of an encoder for the left and right contexts of the entity mention, another encoder for the entity mention itself, and a logistic regression classifier working on the features from the aforementioned encoders.

The above model improves on the work by ~\cite{shimaoka2017neural}. The major drawback in that model is the use of custom hand crafted features, tailored for the specific task, which makes generalization and transferability to other datasets and similar tasks difficult. Building on these ideas, we have attempted to augment neural network based models with low level linguistic features which are obtained cheaply to push overall performance. Below, we elaborate on some of the architectural tweaks we attempt on the base model.

Similar to~\cite{shimaoka2017neural}, we use two separate encoders for the entity mention and the left and right contexts. For the entity mention, we resort to using the average of the word embeddings for each word. For the left and right contexts, we employ the three different encoders mentioned in~\cite{shimaoka2017neural}, viz.

The results on the TACRED dataset as can be seen in Table ~\ref{tab:ontonotes_results}, clearly show the same trend, i.e. adding token level features improve performance across the board, for all metrics, as well as for any choice of encoder. The important thing to note is that these token level features can be obtained cheaply, using off-the-shelf NLP tools to deliver linguistic features such as POS tags, or using existing rule based systems to deliver task or domain specific type tags. This is in contrast to previous work such as~\cite{ling2012fine},~\cite{yogatama2015embedding} and others, who resort to carefully hand crafted features.

\begin{figure*}[htb]
\begin{subfigure}{0.48\textwidth}
    \includegraphics[width=\columnwidth]{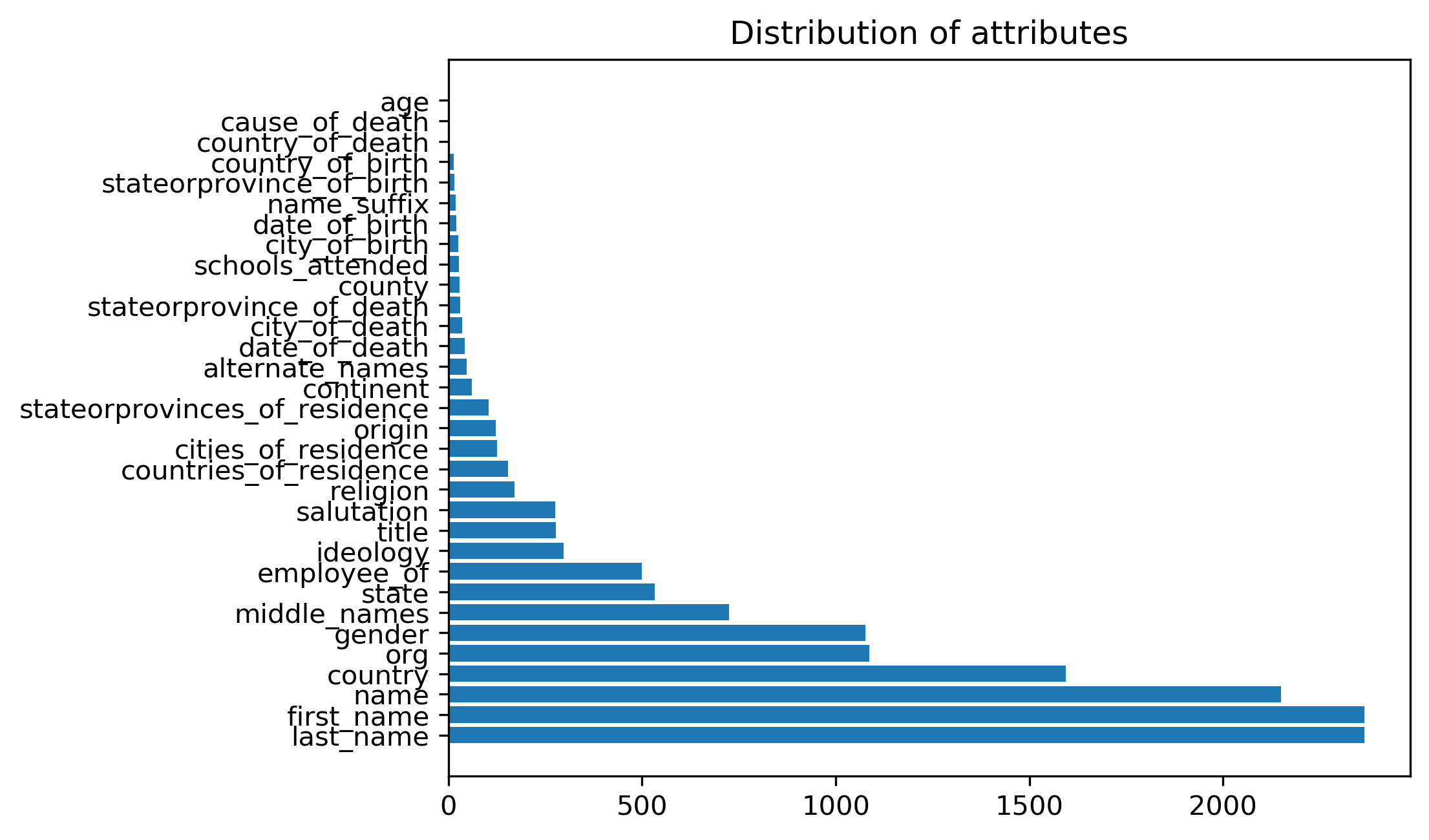}
    \caption{Distribution of node attributes}
    \label{fig:node_attributes_distribution}
\end{subfigure}
\begin{subfigure}{0.48\textwidth}
    \includegraphics[width=\columnwidth]{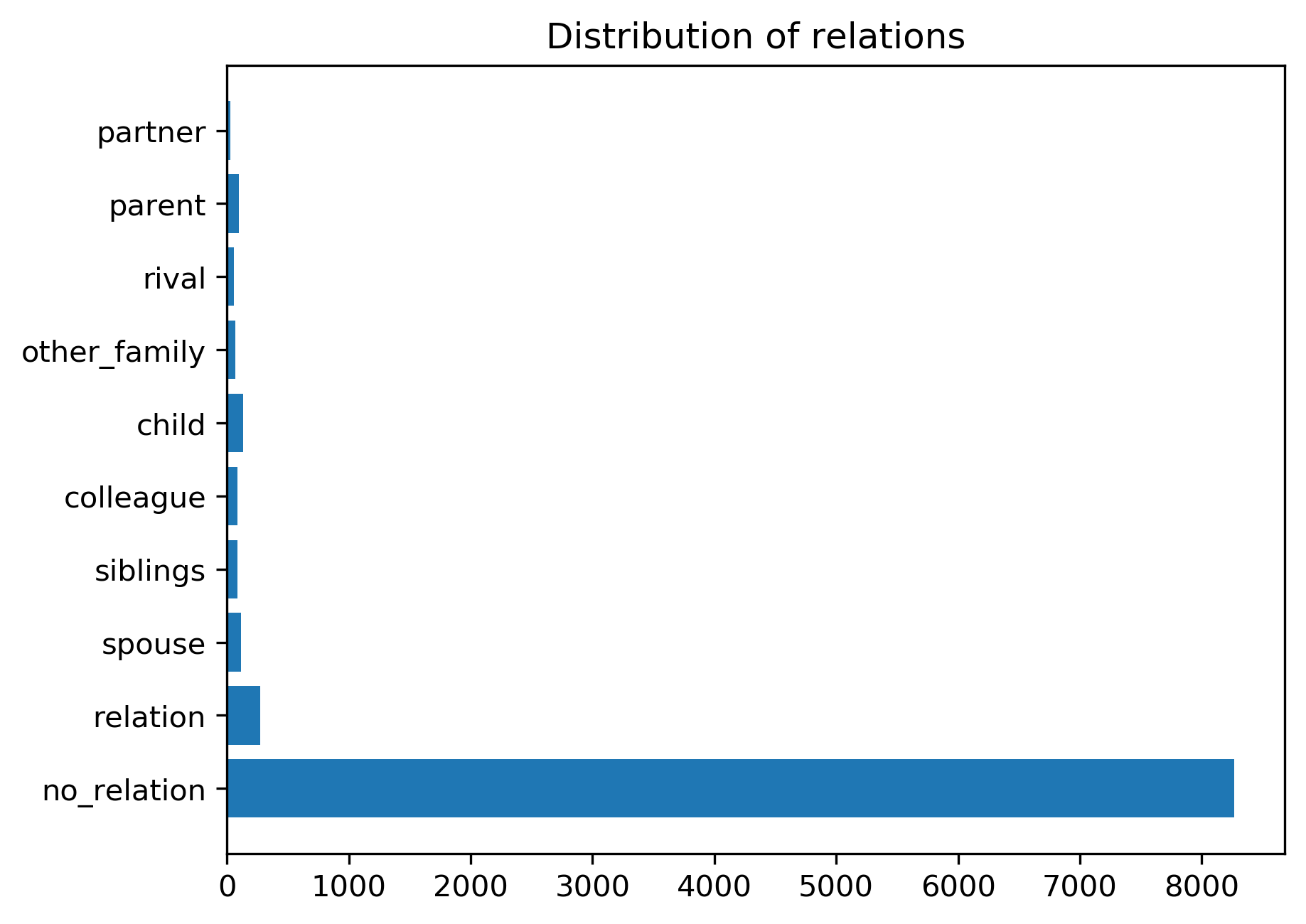}
    \caption{Distribution of relation types}
    \label{fig:relation_distribution}
\end{subfigure}
\end{figure*}

\begin{table}[!htb]
    \begin{center}
    \begin{tabular}{cccc}
    \hline
    \textbf{Dataset} & \textbf{Model} & \textbf{Macro F1} & \textbf{Micro F1} \\
    \hline
    {\textbf{OntoNotes}}
        & NFGEC      & 0.678    & 0.617 \\
        & NFGEC+       & 0.740    & 0.672 \\
        \cline{2-4}
    \hline
    \end{tabular}
    \caption{Entity Classification performance with and without augmented  features.}
    \label{tab:ontonotes_results}
    \end{center}
\end{table}

% Experiments Model
\subsection{Relation Extraction}

\begin{figure}
    \centering
    \includegraphics[width=\columnwidth]{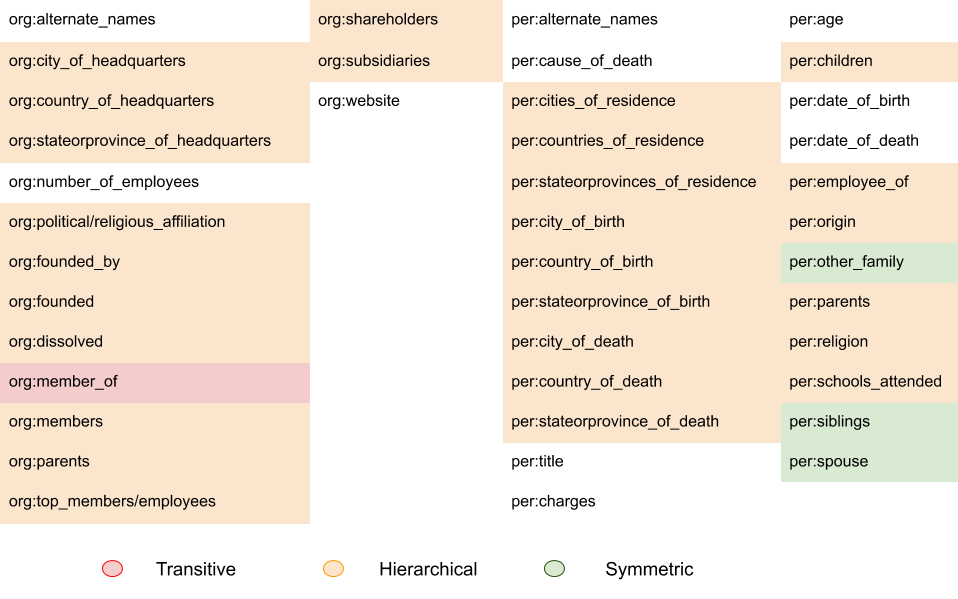}
    \caption{TACRED Relations}
    \label{fig:tacred_relations}
\end{figure}

While extracting entities from unstructured text can be improved as we have shown in Table \ref{tab:ontonotes_results}, extracting people to people relations is a harder problem. As we mentioned earlier, \cite{alt2020tacred} and other works have investigated ways to improve the performance of relation extraction model proposed by \cite{zhang2017position}.

Given our motivation to populate a personal knowledge graph, we focus only on the Data Augmentation strategies and GNN based missing link prediction problem in our experiments. As shown in Table \ref{tab:ontonotes_results} for entity classification, we believe Neural Networks for Relation Extraction can also use features from data augmentation to improve their performance. We however leave this for future work.

Instead, we have augmented the results available in the TACRED Dataset with rule based annotations and the a Link Prediction model. We defer the discussion on the augmentation performance to Section \ref{cold_start}.

% Experiments Model
\subsection{Link Prediction}

As shown in \ref{tab:linkprediction}, Position Aware Graph Neural Network performs better than Graph Convolutional Networks on the TACRED Dataset. We first evaluate the performance of the Link Prediction using TACRED Dataset as the ground truth. The model thus trained can then be used for predicting links not present in Knowledge Base populated using the pipeline so far.

\begin{table}[!htb]
    \begin{center}
    \begin{tabular}{ccccccc}
    \hline
    \textbf{Dataset} & \textbf{Model} & \textbf{ROC AUC} & \textbf{Std. Dev.} \\
    \hline
    {\textbf{TACRED + Augmenation}}  &   \textbf{GCN} & 0.4047  & 0.09184  \\
                    &   \textbf{P-GNN} & 0.6473 & 0.02116   \\
    \hline
    \end{tabular}
    \caption{Link Prediction models performance on TACRED Dataset}
    \label{tab:linkprediction}
    \end{center}
\end{table}

The use of a Link Prediction model instead of a slot filling model is an improvement over traditional Knowledge Base Population methods. However, GNN models are harder to interpret especially in a homogenous property graph set up like ours. Further they could lead to bias on protected variables. Hence we present fairness analysis in Section 6, to ensure that our KBP process is agnostic to protected variables. 

The Personal Knowledge Graph populated by us can be used to improve search, natural language based question answering, and reasoning systems. Further the graph data can be exported to other data formats like the RDF and PPI formats, and used as a dataset for Link Prediction experiments.

% Results    
\section{Cold Start KBP Evaluation}
\label{cold_start}

Similarly as per the previous Research, to evaluate we use the strong basline evaluation method of the TAC KBP 2015 cold start slot filling task \cite{ellis2015overview} to check performance of the models on populating the knowledge graph comparing with tacred model. The entities for hop-0 and hop-1 are manually selected using random search for particular persons present in the corpus data. Corresponding queries are generated for hop-0. The predicted relation data for hop-0 in turn serves as the input entity for the hop-1 slot, for which a particular query related to it is manually generated. Hop-0 generally involves the person to person relation and hop-1 consists of person to attribute relation.

Finally the precision, recall and F1-Score (micro) is calculated for hop-0 and hop-1. The error in hop-0 easily propagates to hop-1 as well. To fairly evaluate each relation extraction model on this task, Stanford's 2015 slot filling technique is used. It is the top ranked evaluation baseline specifically tuned for KBP evaluation. Then later the evaluation metrics for hop-all is developed as the combination of both hop-0 and hop-1 slots.
As shown in \ref{tab:kbpevaluation1} the recall of the TACRED is less as our augmented dataset with recall about 0.52.
 
\begin{table}[!htb]
    \begin{center}
    \begin{tabular}{cccc}
    \hline
    \textbf{Metric} & \textbf{Precision} & \textbf{Recall} & \textbf{F1} \\
    \hline
     {\textbf{hop0}}  &  1.00 & 0.696 & 0.821   \\
    {\textbf{hop1}}  &   1.00 & 0.521  & 0.685  \\
    {\textbf{hop-all}}  &   1.00 & 0.609  & 0.753 \\
    \hline
    \end{tabular}
    \caption{Compared to our augmented dataset, baseline TACRED has less recall.}
    \label{tab:kbpevaluation1}
    \end{center}
\end{table}

Though the augmentation performs ver well, on adding the 

\begin{table}[!htb]
    \begin{center}
    \begin{tabular}{cccc}
    \hline
    \textbf{Metric} & \textbf{Precision} & \textbf{Recall} & \textbf{F1} \\
    \hline
     {\textbf{hop0}}  &  1.00 & 0.737 & 0.849   \\
    {\textbf{hop1}}  &   1.00 & 0.576  & 0.731  \\
    {\textbf{hop-all}}  &   1.00 & 0.657  & 0.791 \\
    \hline
    \end{tabular}
    \caption{Augmention provides most of the recall while GNNs add some relations too.}
    \label{tab:kbpevaluation}
    \end{center}
\end{table}

We also observed that the Recall of tacred for the protected attributes before augmentation is only around 0.095 whereas the recall of our solution for protected attributes after augmentation is around 0.99. For example the religion of Bill Clinton could not be extracted by the tacred model rather extracted by the augmented model. 

\begin{table}[!htb]
    \begin{center}
    \begin{tabular}{cccc}
    \hline
    \textbf{Metric} & \textbf Tacred & \textbf Our Solution \\ 
    \hline
    {\textbf{Precision}} & 1.00 & 1.00    \\
    {\textbf{Recall}}    & 0.095 & 0.972 \\
    {\textbf{F1-score}}  & 0.174 & 0.984 \\
    \hline
    \end{tabular}
    \caption{Our solution has much higher recall on protected variables}
    \label{tab:kbpevaluation}
    \end{center}
\end{table}

\section{Fairness Analysis}

\begin{figure*}[htb]
    \includegraphics[width=\linewidth]{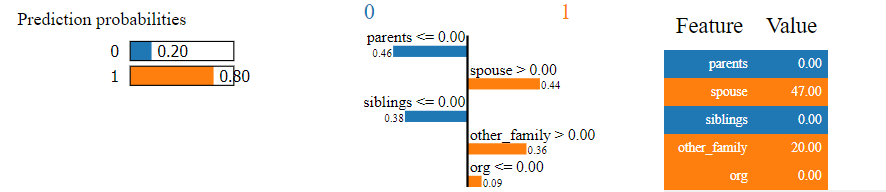}
    \caption{Feature Importance analysis using LIME}
    \label{fig:featureweight_lime}
\end{figure*}

\begin{figure}[htb]
    \includegraphics[width=\columnwidth]{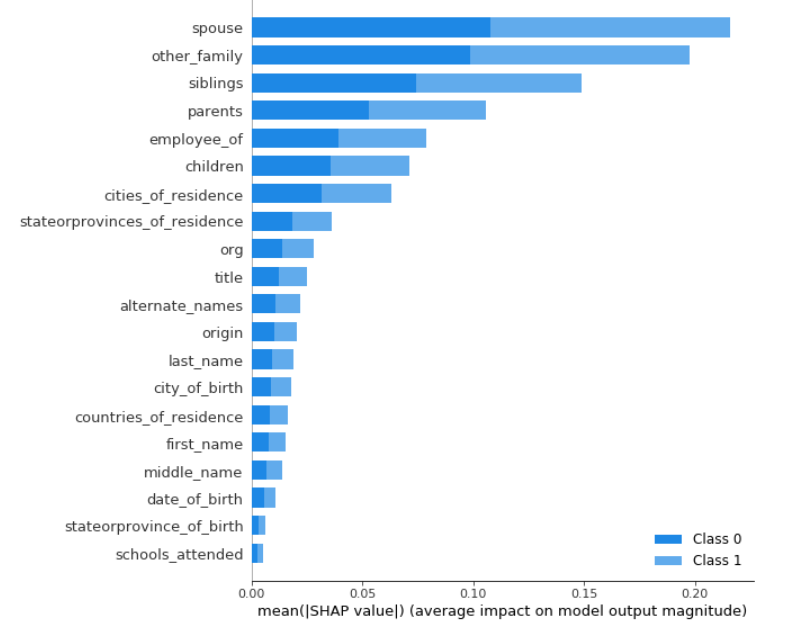}
    \caption{Feature Importance analysis using SHAP values}
    \label{fig:shapvalues}
\end{figure}

  The links created on the Personal Knowledge Graph populated is examined by developing a structured dataset from graph.The structured dataset developed has binary target class(0- one or more persons linked, 1- no relation). Using an interpretable model to analyze a complex model as a post-hoc analysis is currently prevalent in this field of research. The goal here is to see if the relations are dependent on the protected variables.

\begin{table}[!htb]
    \begin{center}
    \begin{tabular}{cccc}
    \hline
    \textbf{Metric} & \textbf{Precision} & \textbf{Recall} & \textbf{F1} \\
    \hline
     {\textbf{no relation}}  &  1.00 & 1.00 & 1.00   \\
    {\textbf{one or More person related}}  &   1.00 & 0.95  & 0.98  \\
    {\textbf{macro avg}}  &   1.00 & 1.00  & 1.00  \\
    {\textbf{weighted avg}}  &   1.00 & 0.98  & 0.99  \\
    \hline
    \end{tabular}
    \caption{Interpretable Random Forest model}
    \label{tab:random_forest_model}
    \end{center}
\end{table}

We train a simple binary classifier on whether a person is a linked to one or more people or not related to any person. The Knowledge Graph populated is converted into tabular structured data for each person in the graph. Using the structured data the presence or absence of a link from a person to any other person is predicted.

After training this classifier to predict links,  features of that binary classifier are analyzed using LIME and SHAP. Logistic Regression, Decision Trees, Random Forest and XGboost are used to model the structured data and predict whether a link is predicted for the person with any other person in the Knowledge Base.The better performing Random Forest model is finally chosen for classification. Table \ref{tab:random_forest_model} represents the classification report of Random Forest model to predict target classes with an F1 score of 90 percent stating that the person belongs to that target class 1 because of he/she is related by their organization, family or siblings.

Table \ref{tab:random_forest_model} shows that the important features like city of birth,residence, family relations like siblings,spouse are considered in the TACRED Model while building the knowledge graph. The features of men and women are extracted equally by models as they do not influence the model classification. Figure \ref{fig:featureweight_lime} shows that the  feature organization,family relations is main attribute to predict whether a new person added to the Knowledge Base is going to be linked to any other person in the graph. Note 1 in \ref{fig:featureweight_lime} shows that the attributes to left does not impact lot to the prediction of target class 1 as compared to the impactful features to the right of vertical bar.

 Figure \ref{fig:shapvalues} shows the graph of SHAP values magnitude of each feature on the resulting output of target class. residence,city of birth is considered as contributing to the target class 1 as those persons are born in same city increasing the chance of relation between them, family relations is highly contributing to  target class 1 for predicting links between two persons. date of birth and city of death are least contribution to state whether a person is linked or not.

\begin{figure*}[htb]
\begin{subfigure}{0.48\textwidth}
    \includegraphics[width=\columnwidth]{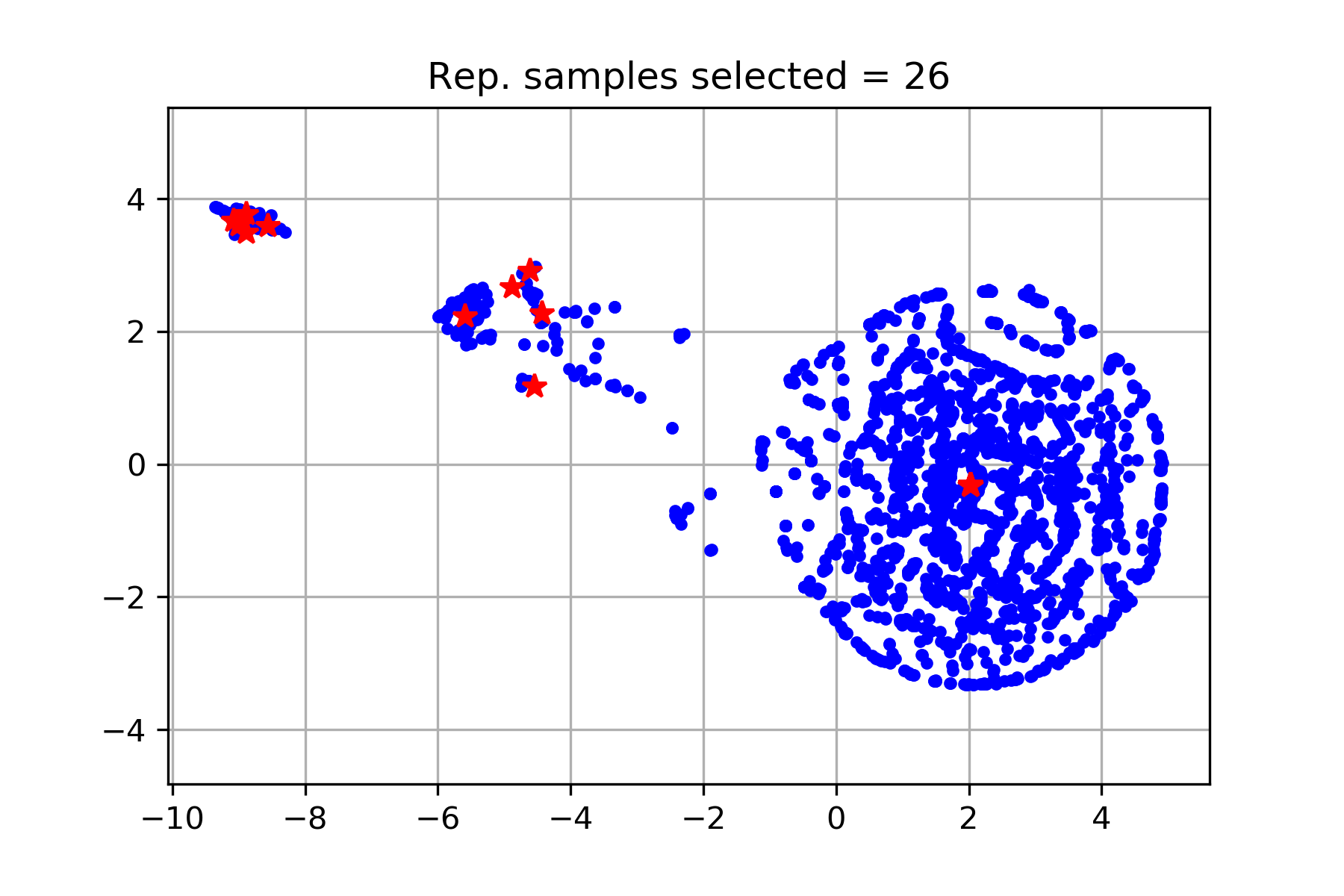}
    \caption{Sampling before augmentation}
    \label{fig:representative_set_sampling}
\end{subfigure}
\begin{subfigure}{0.48\textwidth}
    \includegraphics[width=\columnwidth]{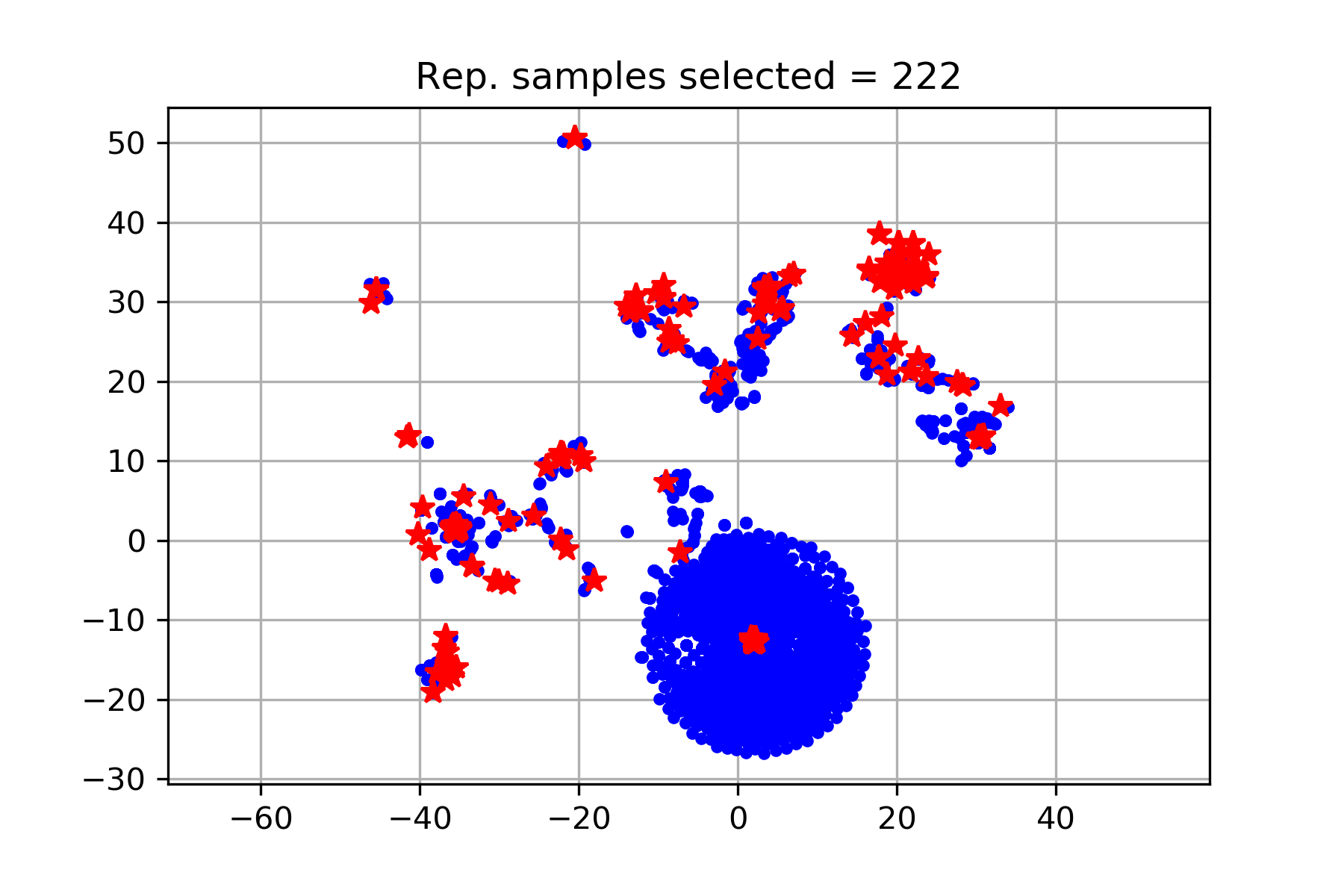}
    \caption{Sampling after augmentation}
    \label{fig:representative_set_sampling2}
\end{subfigure}
\caption{This is the result of running our sampling algorithm on the dataset. The algorithm selected a total of 26 and 222 data points respectively from the baseline TACRED dataset and our augmented dataset. The figure shows TSNE projection of the dataset with blue dots denoting the points in the dataset and red dots denoting the representative samples. The skewness can be explained by the fact that we're drawing only people nodes and their personal attributes from TACRED dataset.}
\end{figure*}

\section{Representative Set Sampling}

In the process of generating the knowledge graph, we might want to manually inspect the generated entities to have an understanding of what kind of entities are generated and how good are the attributes associated with them. Inspecting all the entities is a tiresome job and hence we might want to inspect a small subset of the entities. To get meaningful insights about the dataset, we need the samples that we select to be representatives of the entire dataset. One simple approach to sampling could be random sampling where we select samples at random. There can be other deterministic approaches like select every $k^{th}$ entity in the dataset and inspect them etc. But, there are some issues associated with the simple approaches above.
\begin{itemize}
    \item \textbf{Redundancy} : If we assume that the entities are generated form a probability distribution P, then it is more likely that random sampling would yield samples with patterns around the mean/mode etc. Hence the  samples in the subset would mostly convey redundant information.
    \item \textbf{Coverage} : Because random sampling yields samples with patterns around the mean/mode, we are more likely to ignore data patterns with lesser probabilities in the subset. Hence to have a better coverage, we would want the samples to be representatives of most of the patterns in the dataset.
\end{itemize}
To circumvent the issues above, we develop a sampling algorithm that employs some heuristics to reduce the redundancy and increase coverage of patterns among the samples that are selected. The desiderata for the sampling algorithm are as follows :
\begin{itemize}
    \item Select at least one sample which represents the most frequent pattern in the dataset. This ensures coverage of the dataset.
    \item No two patterns selected in the subset should be similar by more than a threshold $\theta$. This avoids redundant patterns in the subset. 
\end{itemize}

Our sampling algorithm  is inspired from matrix sketching \cite{liberty2012simple}, which is a linear time algorithm to find the most frequent patterns in the dataset. More formally, given a matrix $A \in \mathcal{R}^{n\times m}$, the algorithm finds a smaller sketch matrix $B\in \mathcal{R}^{l\times m}$ such that $l<<n$ and $A^TA \approx B^TB$. Here we can observe that the matrix $B$ tries to capture most of the variance in $A$. In other words, each row of $B$ represents a frequent direction in $A$ and also because $B$ is obtained by performing $SVD$ on rows of $A$, each row of $B$ is orthogonal to other rows of $B$. Our intuition is that, once we get the frequent directions of $A$, we can easily select data points along that direction and thereby select samples representing the frequent patterns in the dataset. The sampling algorithm expects the input to be in numerical form only. We convert each categorical attribute to one hot embedding and normalize each numerical column to be between $[0,1]$ and feed it as input to the algorithm. We drop other text attributes. Hence, input to the sampling algorithm is a matrix $A$ that is scaled for numerical attributes and one-hot embedded for categorical attributes respectively. \\
Now, we explain the sampling algorithm in several steps. \\

\begin{enumerate}
    \item \textbf{Dimensionality reduction} : Going forward in the algorithm, we would apply matrix sketching on the input matrix $A$, which would in turn apply SVD. Hence to keep the eigen problem tractable, we do feature agglomeration to reduce the dimensionality of $A$. In our experiments we retained only the top 100 features of A.
    \item \textbf{Clustering} : One of the popular methods in literature to select representatve points is to partition the dataset into many clusters and select representatives from each of them. We also adapt the same paradigm. We first cluster the dataset and apply sampling algorithm in each cluster independently. Thereby, we select representative points from each cluster. In our experiments, we run KMeans algorithm with number of clusters = $k$, where $k$ is determined by the elbow method.
    \item \textbf{Select data points representing frequent patterns} : For each cluster $i \in [k]$, we select the cluster matrix $A_i$, where $A_i$ contains the rows of the entities that belong to cluster $i$. We then apply matrix sketching algorithm on $A_i$ to find a set of frequent directions $\{v_j\}$ in $A_i$. Then, for each frequent direction $v_j$, we select a data point $a_j \in A_i$ along that direction and include it in the representative subset. 
    \begin{equation}
        a_j = argmax_{a \in A_i} cosine(a, v_j)
    \end{equation}
    \item \textbf{Remove redundant points} : Once we select points along the frequent directions, we remove points that are redundant to the selected points. For each $a_j$ selected in step-3, we collect all the redundant points in $A_i$ as 
    \begin{equation}
        R_j = \{r_j \in A_i | cosine(r_j, a_j) \geq \theta\}
    \end{equation}
    In our experiments, we set $\theta = 0.85$. Once we collect the redundant points $R_j$, we remove them from the cluster. i.e. $A_i = A_i\setminus R_j$
\end{enumerate}
We repeat Steps 3 and 4 above until we exhaust the data points in the cluster $A_i$. Intuitively, it is easy to observe that step 3 selects points representing the frequent points in the dataset and step 4  avoids selecting points representing redundant patterns in the representative subset. \\ The results of the sampling algorithm is shown in Figure \ref{fig:representative_set_sampling}. We manually inspected the representative samples of generated entities and their associated links and observed that the proposed method of personal KG populations yields satisfactory results.

\section{Conclusion}
Using a combination of neural models and rule based annotators known as SystemT, we showed how data augmentation improves the overall diversity of a personal knowledge base populated from the TACRED dataset. We then performed a detailed fairness analysis and representative set sampling to show this personal knowledge base can be used to train Graph Neural Network models for predicting links between people in the real world. Our work seems to show that data augmentation can help train GNNs with reasonable confidence that they are not biased against minority groups.

%\begin{acks}
%This work was done as part of the Global Remote Mentoring initiative of IBM %University Relations to promote undergraduate student research. We thank %Kalapriya Kannan, Dinesh Garg, Poornima Iyengar, Kranti Athalye, The Principal %and Management of NMIT for their support.
%\end{acks}

\bibliographystyle{ACM-Reference-Format}
\bibliography{acmart}

\end{document}